\documentclass[%
 reprint,
 amsmath,amssymb,
pra,
]{revtex4-2}

\usepackage{graphicx}% Include figure files
\usepackage{dcolumn}% Align table columns on decimal point
\usepackage{multirow}
\usepackage{amsmath}
\usepackage{amssymb}
\usepackage{euscript}
\usepackage{bm}% bold math
\usepackage{xcolor}
\usepackage{lipsum}
\usepackage{hyperref}
\usepackage{qcircuit}
\usepackage{graphicx}% Include figure files
\graphicspath{{figures/}}
\usepackage{dcolumn}% Align table columns on decimal point
\usepackage{bm}% bold math
\usepackage{bbold}
\usepackage[utf8]{inputenc}
\usepackage[T1]{fontenc}
\usepackage{mathptmx}
\usepackage{mathtools}
\usepackage{xcolor}
\usepackage{tikz}
\usepackage{braket}
\usepackage{wrapfig}

\definecolor{ao(english)}{rgb}{0.0, 0.5, 0.0}

\usepackage[english]{babel}
\newtheorem{theorem}{Theorem}

\begin{document}

\preprint{AIP/123-QED}

\title[TDHF-YB]{Hybrid algorithm for the time-dependent Hartree--Fock method using the Yang--Baxter equation on quantum computers}

% Force line breaks with \\

\author{Sahil Gulania}
\email{sgulania@anl.gov}
\affiliation{Mathematics and Computer Science Division, Argonne National Laboratory, Lemont, Illinois 60439, United States}

\author{Stephen K. Gray}
\email{gray@anl.gov}
\affiliation{Center for Nanoscale Materials, Argonne National Laboratory, Lemont, Illinois 60439, United States}

\author{Yuri Alexeev}
\email{yuri@anl.gov}
\affiliation{Computational Science Division, Argonne National Laboratory, Lemont, Illinois 60439, United States}

\author{Bo Peng}
\email{peng398@pnnl.gov}
\affiliation{Physical and Computational Sciences Directorate, Pacific Northwest National Laboratory, Richland, Washington 99352, United States}

\author{Niranjan Govind}
\email{niri.govind@pnnl.gov}
\affiliation{Physical and Computational Sciences Directorate, Pacific Northwest National Laboratory, Richland, Washington 99352, United States}

\date{\today}
\begin{abstract}
 The time-dependent Hartree--Fock (TDHF) method is an approach to simulate the mean field dynamics of electrons within the assumption that the electrons move independently in their self-consistent average field and within the space of single Slater determinants.
%behavior of the electronic density under
%the presence of a time-dependent electric field. 
One of the major advantages of performing time
dynamics within Hartree--Fock theory is the free fermionic nature of the problem, which makes TDHF classically simulatable in polynomial time. Here, we present a hybrid TDHF implementation     
for quantum computers. This quantum circuit grows with time; but with our recent work on circuit compression 
via the Yang--Baxter equation (YBE), the resulting circuit is constant depth. 
This study provides a new way to simulate TDHF with the aid of a quantum device as well as provides a new direction for the application of YBE symmetry in quantum chemistry simulations. 
%\textcolor{red}{SKG: need to state signficance of the circuit approach given that we state TDHF scales in polynomial time on classical computers.}
\end{abstract}
\maketitle

%\section{TO DO List}
%\begin{enumerate}
%    \item Introduction - Niri and Stephen and Bo
%    \item Theory - Bo
%    \item Proof and Result - Sahil 
%    \item Conclusion - Yuri
%\end{enumerate}

%\section{Introduction}
%\begin{enumerate}
%    \item 
%    Alternative way of doing TDHF on Quantum Computers which bypass the diagonalization at each step
    
%    \item Compare it with classically method
%    \item Explain the importance of compression
%    \item explain the relation with matchgates 
%\end{enumerate}

%\newpage

The time-dependent Hartree--Fock~\cite{Going2018,Li2020,jorgensen1975molecular}(TDHF) method is a widely used mean field approach to study the dynamics of many-electron systems. 
It is based on the  Hartree--Fock~\cite{hartree28,Fock30,Whitfield13b,Google20} (HF) theory, which 
\iffalse
which is a fundamental approach in quantum mechanics used to approximate the behavior of many-body systems, particularly those involving electrons in atoms, molecules, and solids. It 
\fi
provides a self-consistent field method for describing the electronic structure of these systems by treating the electrons as indistinguishable particles subject to the Pauli exclusion principle~\cite{kaplan2017pauli,klyachko2009pauli}.
The Hartree-Fock equations involve solving an eigenvalue problem, where the Fock operator, representing the effective mean field potential experienced by each electron due to the other electrons, acts on the spin orbitals. This operator takes into account the Coulomb repulsion between electrons and the exchange interaction resulting from their indistinguishability. TDHF extends static HF to the time domain by incorporating the relevant time dependence either perturbatively or non-perturbatively.\cite{Li2020}
\iffalse
The TDHF method extends the HF 
the static HF method by incorporating time-dependent perturbations, enabling the calculation of the time evolution of the electronic wave function and providing insight into the dynamics of quantum systems.
\fi
Even though the motions of the electrons are not completely independent, TDHF is typically defined as uncorrelated and is insufficient for the treatment of strongly correlated systems. Nevertheless, it provides an essential baseline for the development of more advanced time-dependent approaches such as time-dependent density functional theory~\cite{Lopata2012,Zhao2020,Li2020}, time-dependent configuration interaction, ~\cite{greenman2010,krause2007molecular} and other time dependent formulations~\cite{Li2020}.

%Despite its importance, TDHF has some limitations. One major limitation is its computational demand, as it can be difficult to implement for systems with many electrons, which poses challenges for large-scale simulations. Furthermore, the HF approximation upon which TDHF is based can lead to inaccuracies in the results, particularly in systems with strong electron-electron interactions or highly correlated systems.

In recent years, quantum computing ~\cite{DOE-QIS,preskill2018quantum} has emerged as a promising platform to tackle challenging problems in various scientific domains, including quantum chemistry. Although there is no direct quantum advantage~\cite{jones2022chemistry} for TDHF simulations using quantum computing, implementing TDHF on a quantum computer can offer several benefits. It can pave the way for the development of a general framework for simulating quantum time dynamics, which could be applicable to other time-dependent quantum chemistry methods or even other quantum systems. Furthermore, TDHF simulations on quantum computers can serve as a valuable benchmark for validating and comparing the performance of quantum algorithms, error correction techniques, and hardware.
By developing and implementing TDHF simulations on quantum computers, we can prepare for future quantum advantages that may arise from more advanced quantum algorithms or the discovery of new problems in quantum chemistry.  Moreover,  implementing TDHF simulations on quantum computers helps train a new generation of researchers in the principles and techniques of quantum computing while exposing them to the challenges and opportunities of applying quantum computing to real-world problems in quantum chemistry.

In this paper we explore the application of quantum computing to TDHF simulations, focusing on circuit optimization for free fermions making use
of the Yang--Baxter equation ~\cite{yang1967some,baxter1972partition} of field theory. The approach is hybrid in nature, involving both quantum and classical computers.  We demonstrate in simulation how these circuits can also be used for real-time TDHF dynamics on quantum computers.
%providing a foundation for further research into the application of quantum computing for quantum time dynamics, even beyond the scope of TDHF. 
This interdisciplinary collaboration between quantum chemistry, quantum field theory, and quantum computing can lead to novel insights.
%and innovative solutions. 
This work also serves as a stepping stone for developing more advanced and efficient quantum algorithms for simulating time-dependent quantum systems, thereby broadening the potential applications of quantum computing in quantum chemistry and related fields.

First, we introduce the TDHF method and Yang--Baxter equation. We then explain the construction of a quantum circuit for TDHF and how the Yang--Baxter relation mediates the compression of the quantum circuit. We also lay out the step-by-step procedure for performing the quantum time dynamics using a hybrid classical-quantum algorithm.
We then discuss the simulation of a hydrogen molecule being driven by a time-dependent electric field on a quantum simulator. We conclude  with a summary. 

\section{Theory}

%%%%%%%%%%%%%%%%%%%%%%%%%%%%%%%%%%%%%%%%%%%%%%%%%%%%%%%%%%%%%%%%%

%\subsection{Time-dependent Schr\"{o}dinger equation (TDSE)}
\subsection{Time-dependent Hartree--Fock method}
%In TDHF, the time evolution of the electronic wave function is described by 
The time-dependent Schrödinger equation (TDSE) for the evolution of the wavefunction or state vector $| \psi (t) \rangle$ is
%~\cite{tal1984accurate,leforestier1991comparison,mclachlan1964variational} (TDSE) is a fundamental equation in quantum mechanics that describes how the wave function of a quantum system evolves in time. It provides a mathematical framework to study the time evolution of quantum states and predict the behavior of quantum systems. For an $N$-electron quantum system characterized by a Hamiltonian $\hat{H}$, its TDSE is a first-order linear differential equation that describes how the wave function of the $N$-electron quantum system, $|\psi(t)\rangle$, changes over time $t$, i.e.,
%
\begin{equation}
    i\frac{\partial |\psi(t)\rangle }{\partial t} = \hat{H}(t)|\psi(t)\rangle,
\label{tdse}
\end{equation}
where $\hat{H}(t)$ could represent the Hamiltonian operator for an $N$-electron system being driven by an external time-dependent
field. Throughout we assume atomic units such that $\hbar$ = 1.
%For a closed, isolated system $\hat{H}$ is time-independent, and only becomes time-dependent when including interaction between the system and time-dependent external forces. 
%Suppose one has the state vector at time $t$, |\psi(t)\rangle$. The solution of Eq. \ref{tdse} at time $t + \delta t$ can be written as
%
%\begin{align}
%    |\psi(t+\delta t)\rangle = \hat{U}(t+\delta t,t) |\psi(t)\rangle
%\end{align}
%
%where  $\hat{U}(t+\delta t,t)$ is the propagator.  In the limit of small $\delta t$ we can write
%
%\begin{align}
%   \hat{U}(t+\delta t,t) \approx e^{-i\hat{H}(t)\delta t}
%\end{align}
Owing to the size of the underlying Hilbert space, solving the TDSE can be challenging for complex systems, and often approximate numerical methods are employed. 
The time-dependent Hartree--Fock (TDHF) method \cite{Going2018,Li2020,jorgensen1975molecular} is an extension of the traditional time-independent Hartree--Fock method to describe the time evolution of quantum systems. While time-independent Hartree--Fock provides a static description of electronic structure, 
TDHF allows for the investigation of various time-dependent phenomena, such as electronic excitations, chemical reactions, and optical processes. 
%It provides insights into the dynamics of electronic states and their evolution under external perturbations, such as electromagnetic fields or time-dependent potentials.
In TDHF, the wavefunction is expressed as a time-dependent Slater determinant, similar to the time-independent case. Instead of focusing on the stationary Schrödinger equation, however, TDHF approximates the solution of the TDSE, incorporating the time dependence explicitly.
The TDHF equations are derived by extending the stationary Hartree--Fock equations to account for the time dependence of the wavefunction. The equations describe the time evolution of the occupied molecular orbitals (MOs) and their corresponding occupation numbers.
TDHF approximately solves Eq.~\ref{tdse} using the HF approximation for $N$-electrons, in which $|\psi(t)\rangle$ is expressed as a Slater determinant of the time-dependent single-particle orbitals,
\begin{equation}
    |\psi(t)\rangle = \ket{\phi_1(t)\phi_2(t)\cdots\phi_N(t)},
\end{equation}
where $|\phi_i(t)\rangle$ are one-electron molecular orbitals, which can be written as a linear combination of one-electron atomic orbitals ($\chi_{k}$) as
\begin{equation}
    \ket{\phi_{i}(t)} = \sum_{k}^{M} C_{ki}(t) \ket{\chi_{k}},
\end{equation}
where $C_{ki}$ are the coefficients for $M$-atomic orbitals. We can use this to write the Fock matrix as
\begin{equation}
\begin{split}
    &F_{pq}(t) = h_{pq} +  \\
    &\sum_{k}^{M}\sum_{rs}^{M} C_{rk}^{*}(t) C_{sk}(t) (\braket{\chi_{p}\chi_{s}|\frac{1}{r_{ee}}|\chi_{q}\chi_{s}}-\braket{\chi_{p}\chi_{r}|\frac{1}{r_{ee}}|\chi_{s}\chi_{q}}),  
\end{split}
\end{equation}
where $h_{pq}$ 
represents the sum of electron kinetic energy and nuclear-electron attraction terms,
\begin{equation}
    h_{pq} = \braket{\chi_{p}|-\frac{1}{2}\nabla^{2}-\sum_{i}^{nuc} \frac{Z_{i}}{r_{ei}}|\chi_{q}}~~,
\end{equation}
$r_{ee}$ is the distance vector between two electrons, and $r_{ei}$
is the distance vector between electron and nucleus $i$. 
One can also define the time-dependent density matrix as 
\begin{equation}
    \rho_{rs}(t) = \sum_{k}
    C_{rk}^{*}(t)
    C_{sk}(t) = \ket{\phi_{r}}\bra{\phi_{s}}~~;
\end{equation}
and, using this equation, we can also write the time-dependent Fock matrix using the density matrix as
\begin{equation}
\begin{split}
    & F_{pq}(t) = h_{pq} + \\ 
    &\sum_{rs} \rho_{rs}(t) (\braket{\chi_{p}\chi_{s}|\frac{1}{r_{ee}}|\chi_{q}\chi_{s}}-\braket{\chi_{p}\chi_{r}|\frac{1}{r_{ee}}|\chi_{s}\chi_{q}}) .
\end{split}
\end{equation}
%
%\textcolor{blue}{BP: what is subscript "i" referring to? How is your Hamiltonian defined? the above equation is abusing symbols, as it doesn't distinguish the system Hamiltonian from the single-particle Hamiltonian.}
The time-dependent mean-field Hamiltonian $\hat{h}(t)$ has two terms: (1) the mean-field interaction of electrons at time $t$ from the Fock operator and (2) the interaction of electrons with the field at time $t$. 
The interaction with field can be computed by using the dot product between the dipole moment matrix ($\hat{D}$) interacting with field $E(t)$ at a given time $t$. Therefore, the total mean-field Hamiltonian becomes
\begin{equation}
\hat{h}(t) = \hat{F}(t) + \hat{D} \cdot E(t),
\end{equation}
%\textcolor{blue}{BP: what is the subscript 'b' in J and K referring to? How are J and K defined.}.
and the equations for the molecular orbitals $\phi_{i}(t)$ can be written in
terms of the Fock operator as
\begin{equation}
    i\frac{\partial |\phi_{i}(t)\rangle}{\partial t} = \hat{h}(t) |\phi_{i}(t)\rangle .
\end{equation}
%\textcolor{blue}{BP: the above equations are quite confusing and not right. The time dependence is not described properly. BTW, what is 'm' in Eq. (9)? what is 'D' in Eq. (10)? why is it time-dependent?}
The  evolution of the full wavefunction is 
\begin{equation}
   | \psi(t+\delta t ) \rangle = \hat{U}(t+\delta t ,t) | \psi(t) \rangle ,
\end{equation}
where, valid for small $\delta t$, the approximation
\begin{equation}
   \hat{ U}(t+\delta t, t) \approx e^{-i \hat{h}(t) \delta t}
\end{equation}
is made.
After each time step, one must update the $\hat{h}(t)$ in order to take the next time step. 
Thus, assuming one has advanced to $| \psi (t + \delta t ) \rangle$  as above, one needs
to construct $\hat{h} (t+ \delta t)$ in order to take the next step to time $t + 2 \delta t$
This can be achieved by computing the density matrix at time $t+\delta t$: 
\begin{equation}\label{eq:density}
    \rho(t+\delta t) = \ket{\psi(t+\delta t)} \bra{\psi(t+\delta t)}~~.
\end{equation}
This density matrix is used to compute $\hat{F}(t+\delta t)$, which is further used to update $\hat{h}(t+\delta t)$ and to propagate the wavefunction to $| \psi(t+ 2 \delta t) \rangle$. 
%\textcolor{blue}{BP: better to show the equations. That is how real-time TDHF works.}. 
%This numerical process is followed due to coupled differential equation, which does not have analytical solution. 
Given that $\delta t$ is small enough, the numerical solution approaches the exact solution of the  TDHF equations. We will  show later  how to compute the density matrix both in the molecular basis and in the atomic basis that are used to represent the wavefunction. 
\subsection{Yang--Baxter equation}
The Yang--Baxter equation  was introduced independently in theoretical physics by Yang~\cite{yang1967some} in the late 1960s and by Baxter \cite{baxter1972partition} in statistical mechanics in the early 1970s. This relation has also received much attention in many areas of theoretical physics, classification of knots, scattering of subatomic particles, nuclear magnetic resonance, and ultracold atoms and, more recently, in quantum information science \cite{ge2016YBE,nayak2008non,kauffman2010topological,zhang2013integrable,vind2016experimental,batchelor2016yang}. 
%why introduce YBE in quantum computation

The YBE connection to quantum computing originates from investigating  the relationship between topological entanglement, quantum entanglement, and quantum computational universality. Of particular interest is  how the global topological relationship in spaces (e.g., knotting and linking) corresponds to the entangled quantum states and how the CNOT gate, for instance, can in turn be replaced by another unitary gate $R$ to maintain universality. It turns out that these unitary $\mathcal{R}$ gates, which serve to maintain the universality of quantum computation and also serve as solutions for the condition of topological braiding, are unitary solutions to the YBE \cite{baxter2016exactly}
Briefly, the relation is a consistency or exchange condition that allows one to factorize the interactions of three bodies into a sequence of pairwise interactions under certain conditions. Formally, this can be written as 
\begin{equation}
    (\mathcal{R}\otimes\mathbb{1})(\mathbb{1}\otimes \mathcal{R})(\mathcal{R}\otimes\mathbb{1}) = (\mathbb{1}\otimes \mathcal{R})(\mathcal{R}\otimes\mathbb{1})(\mathbb{1}\otimes \mathcal{R}),
\end{equation}
where the $\mathcal{R}$ operator is a linear mapping $\mathcal{R}: V\otimes V \rightarrow V\otimes V$ defined as a twofold tensor product generalizing the permutation of vector space $V$. This relation also yields a sufficiency condition for quantum integrability in one-dimensional quantum systems and provides a systematic approach to construct integrable models. 
%In vertex models in classical statistical mechanics, the YBE manifests %as a condition on the vertex weights, which allows for the exact %solution of the corresponding model. 
Since a detailed discussion of this topic is beyond the scope of this paper, we refer the reader to more comprehensive works and reviews on the subject \cite{jimbo1989ybe,baxter2016exactly}. 
In the current study, we utilize the similarity between the unitary $\mathcal{R}$ gate and the time propagator for free fermions that we observed in our previous work~\cite{pengYBE2023} to study an efficient quantum simulation of TDHF for general molecular systems.
%\textcolor{red}{SKG: In general the I'm confused on several issues: (a) The YBE, eq 11, is not an identity, it has to be shown to be true for particular problems, right? (b) How is it obvious it is satisfied for the TDHF circuit -- I know this is being demonstrated, I think, below, but it is still not clear to me from all that discussion. (c) If somehow TDHF is then ``integrable'' what is its solution and why do people do it? (d) Would YBE be useful for just a regular quantum chemistry problem cast into quantum computing form via Jordan-Wigner?}
The significance of the Yang--Baxter equation in integrable systems lies in its connection to the concept of ``quantum integrability." Quantum integrability~\cite{caux2011remarks,doikou2010introduction,weigert1992problem,retore2022introduction} refers to the property of a system that allows for the existence of an extensive set of conserved quantities that mutually commute.
The conserved quantities associated with integrable systems are in involution, meaning that their commutators vanish. The Yang--Baxter equation provides a way to construct these conserved quantities and guarantees their mutual compatibility. By solving the Yang--Baxter equation, one can identify the operators that generate the conserved quantities of the system.
We will show later how TDHF quantum circuits follow the YBE and how that allows one to generate a compressed circuit. 
%\begin{enumerate}
%    \item YBE is a condition which when satisfied confirms the integrability
%    \item We need to define integrability and its different form. I am reading about them now. 
%    \item Quantum time dynamics operators show YBE like feature which allows to perform the compression on the circuit. 
%\end{enumerate}

  \section{Circuit Design and Compression}
In this section we show the generalized circuit for time evolution of the mean-field electronic Hamiltonian. We  utilize the YBE to simplify the generalized quantum circuit for an arbitrary time step. First we show how to map the Hamiltonian  operator to Pauli operators. Next we show the merge identity for two-qubit gates. We will also show how reflection symmetry combined with the merge identity allows for the compression of a quantum circuit of any length to a finite depth.
%\textcolor{red}{SKG:This section is titled ''PROOF'' but it is unclear what is being proved.}
In second quantization the Fock operator can be written as 
\begin{equation}
    \hat{F} = \sum_{ij} F_{ij}\hat{a}^{\dagger}_{i}\hat{a}_{j} ,
\end{equation}
where $\hat{a_i} (\hat{a_i}^{\dagger})$ are fermionic creation (annihilation) operators and
\begin{equation}
    F_{ij} = \braket{\phi_{i}|\hat{F}|\phi_{j}} ,
\end{equation}
respectively. Similarly the transition dipole moment can be written in second quantization as
\begin{equation}
    \hat{D} = \sum_{ij} D_{ij} \hat{a}^{\dagger}_{i}\hat{a}_{j} ,
\end{equation}
where 
\begin{equation}
    D_{ij} = \braket{\phi_{i}|\hat{r}|\phi_{j}}.
\end{equation}
Using the Jordan--Wigner representation~\cite{tranter2018comparison}, one can convert any pair of creation and annihilation operators into a Pauli string. Let us consider combinations of the nearest-neighbor pairs only, which are $\hat{a}^{\dagger}_{i}\hat{a}_{i}$, 
$\hat{a}^{\dagger}_{i+1}\hat{a}_{i}$, and 
$\hat{a}^{\dagger}_{i}\hat{a}_{i+1}$. This allows us to convert any local  pair interaction into Pauli operators as
\begin{align}
    \hat{a}_{i}^{\dagger}\hat{a}_{i} &= (I\otimes I-Z\otimes I)/2 \\
    \hat{a}_{i+1}^{\dagger}\hat{a}_{i+1} &= (I\otimes I-I\otimes Z)/2 \\
    \hat{a}_{i}^{\dagger}\hat{a}_{i+1} &= [X\otimes X+Y\otimes Y+i(X\otimes Y-Y\otimes X)]/4\\
    \hat{a}_{i+1}^{\dagger}\hat{a}_{i} &= [X\otimes X+Y\otimes Y+i(Y\otimes X-X\otimes Y)]/4 .
\end{align}
One can also convert this to a matrix representation, which allows us to make their connection to matchgates.
\begin{eqnarray}
a_{i}^\dag a_{i} &=&
    \begin{bmatrix}
    0 & 0 & 0 & 0 \\
    0 & 0 & 0 & 0 \\
    0 & 0 & 1 & 0 \\
    0 & 0 & 0 & 1 
    \end{bmatrix}=\begin{bmatrix}
    0 & 0\\
    0 & 1
    \end{bmatrix}\otimes \begin{bmatrix}
    1 & 0\\
    0 & 1
    \end{bmatrix}\\
    a_{i+1}^\dag a_{i+1} &=&
    \begin{bmatrix}
    0 & 0 & 0 & 0 \\
    0 & 1 & 0 & 0 \\
    0 & 0 & 0 & 0 \\
    0 & 0 & 0 & 1 
    \end{bmatrix}=\begin{bmatrix}
    1 & 0\\
    0 & 1
    \end{bmatrix}\otimes \begin{bmatrix}
    0 & 0\\
    0 & 1
    \end{bmatrix}\\
    a_{i}^{\dagger}a_{i+1} &=& 
    \begin{bmatrix}
    0 & 0 & 0 & 0 \\
    0 & 0 & 0 & 0 \\
    0 & 1 & 0 & 0 \\
    0 & 0 & 0 & 0 
    \end{bmatrix}\\
    a_{i+1}^{\dagger}a_{i} &=& 
    \begin{bmatrix}
    0 & 0 & 0 & 0 \\
    0 & 0 & 1 & 0 \\
    0 & 0 & 0 & 0 \\
    0 & 0 & 0 & 0 
    \end{bmatrix}
\end{eqnarray}
Therefore, concentrating only on a given pair of 
orbitals, one can write the contributions from the Fock and dipole-field interaction terms as
\begin{equation}
   \hat{h}(t) = \sum_{i,i+1}[F_{ij}(t)+D_{ij}~E(t)]\hat{a}^{\dagger}_{i} \hat{a}_{j} .
\end{equation}
To derive the quantum circuit for TDHF, we break down the total mean-field Hamiltonian into parts and construct a circuit for each part. This will allow us to understand the physical importance of each component as well as guide us to derive the total quantum circuit. 

The total Hartree--Fock  mean-field Hamiltonian with dipole interaction involves two parts: a diagonal component, which is simply the occupation numbers, and an off-diagonal part, which represents population transfer from one orbital to another. Therefore, 
\begin{equation}
    \hat{h} = \hat{h}_1 + \hat{h}_2 ,
\end{equation}
where 
\begin{equation}
    \begin{aligned}
    \hat{h}_1 &= \sum_{i} G_{ii}\hat{n}_{i} \\
    G_{ij} &= F_{ij}(t)+D_{ij}~E(t)\\
    \hat{n}_{i} &=\hat{a}^{\dagger}_{i}\hat{a}_{i}
    \end{aligned}
\end{equation}
and 
\begin{equation}
    \hat{h}_2 = \sum_{ij}^{i\neq j} G_{ij}\hat{a}^{\dagger}_{i}\hat{a}_{j} .
\end{equation}
For a given pair of orbitals that are nearest neighbors, the action of time dynamics from $\hat{h}_2$ can be computed as
\begin{equation}
e^{-it \hat{h}_{2}} = e^{-it(G_{i,i+1}a_{i}^{\dagger}a_{i+1}+G_{i+1,i}a_{i+1}^{\dagger}a_{i})} .
\end{equation}
From the Hermitian property of the Hamiltonian, $G_{i,i+1} = G_{i+1,i}$ and using Eqs. (16) and (17), we have
\begin{equation}
    e^{-it \hat{h}_{2}} = e^{-it G_{i,i+1}(X\otimes X+Y\otimes Y)}.
\end{equation}
Similarly, for the given pair the action of time dynamics from $\hat{h}_1$ can be written as 
\begin{equation}
e^{-it \hat{h}_{1}} = e^{-it(G_{i,i}\hat{n}_{i}+G_{i+1,i+1}\hat{n}_{i_1})} ;
\end{equation}
and using Eqs. 20 and 21, we get
\begin{equation}
    e^{-it \hat{h}_{1}} = 
    e^{-it [G_{i,i} (Z\otimes I)+G_{i+1,i+1}(I\otimes Z)]} .
\end{equation}
%\textcolor{red}{SKG: in above firm up notation -- should Fij be F0ij and is it time-dep or time-indep?}
%\textcolor{blue}{SG: Fij is time independent starting point}

\subsection{YBE}
In this section we introduce the Yang--Baxter equation and how it mediates the circuit compression. We then provide a few examples of gate sets that follow YBE. Next we explain the decomposition of the TDHF equations into gates sets and provide a proof that they also follow the YBE relation. 

The Yang--Baxter equation is  well known in condensed-matter physics where it is used to prove the integrability of lattice models. But it is connected to quantum computing from many angles, such as topological entanglement, quantum entanglement, and quantum computational universality. Recently, we showed a new aspect of this equation in the quantum time dynamics of lattice models. Within the Trotter approximation, this equation allows one to perform compression of an arbitrary time step to a finite depth circuit without losing any accuracy. A diagrammatic representation of YBE in the gate set and quantum circuit form is shown in Figure 1.
\begin{figure}[h!]
    \centering
    \includegraphics[width=\columnwidth]{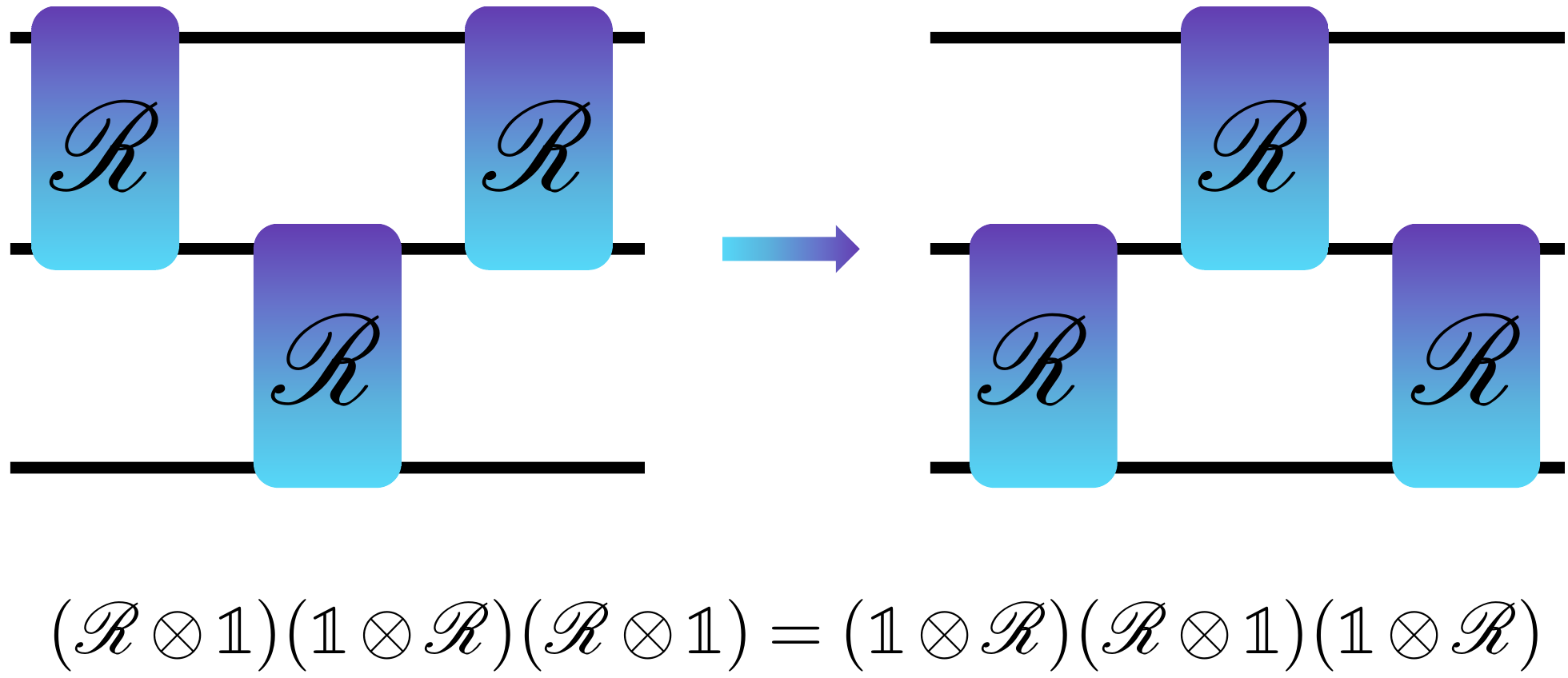}
    \caption{Quantum circuit representation of the YBE for three
qubits.}
    \label{fig:YBE_rep}
\end{figure}
Due to this transformation, one can derive an extra symmetry in the circuit, which is known as mirror symmetry ~\cite{pengYBE2023}. Also, the two-qubit operations follow the merge identity, 
\begin{equation}
    \mathcal{R}_{1}(\theta_1,\theta_2)\times \mathcal{R}_{2}(\alpha_1,\alpha_2) = \mathcal{R}^{'}(\theta_1+\alpha_1,\theta_2+\alpha_2),
\end{equation}
which diagrammatically can be visualized as shown in Figure 2.
\begin{figure}[h!]
    \centering
    \includegraphics[scale=0.25]{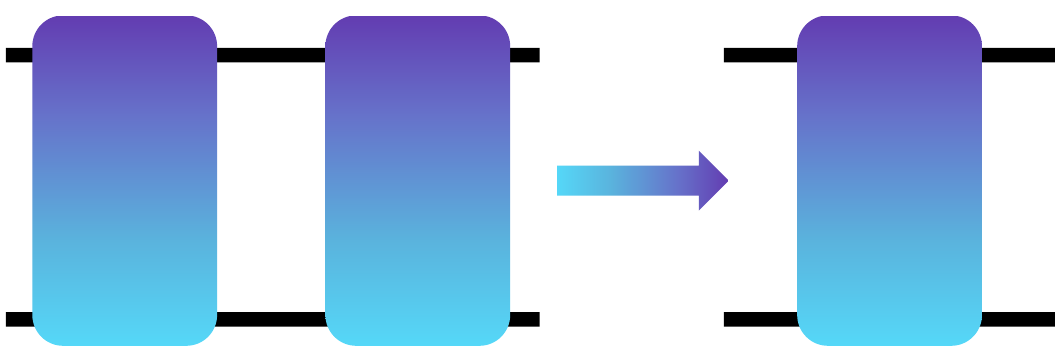}
    \caption{Quantum circuit representation of the merge idenity for two qubits.}
    \label{fig:merge_rep}
\end{figure}
Using mirror identity 
%GAIL - I know there is miror identity, but you just used the term mirror symmetry
and merge identity in a sequential manner, one arrives at a quantum circuit that is constant in depth and scales linearly with system size. The next step  is to make sure all the fermions are able to interact, which can be accomplished using the fermionic swap operator~\cite{kivlichan2018quantum,o2019generalized,hagge2020optimal}.
The fermionic swap operator, also known as the Jordan--Wigner swap operator, is an operator used in quantum physics to exchange the positions of two fermionic modes or particles. It is a generalization of the Pauli spin exchange operator for fermions.
The operator swaps the occupation states of the two fermionic modes and introduces a phase factor. This phase factor arises from  the anti-commutation relations of the fermionic creation and annihilation operators.
\begin{equation}
    \mbox{FSWAP}(\theta) = 
    \begin{bmatrix}
    1 & 0 & 0 & 0 \\
    0 & 0 & e^{i\theta} & 0 \\
    0 & e^{i\theta} & 0 & 0 \\
    0 & 0 & 0 & 1 
    \end{bmatrix}
\end{equation}
The standard QASM~\cite{cross2017open} decomposed quantum circuit implementation of FSWAP comprises three CX gates and a
single-qubit Z rotation.
However, one can use the optimized FSWAP, which is equivalent to the iSWAP operator~\cite{hashim2022optimized}. This operator not only performs the fermionic swap with opposite phases but also follows the YBE; see Figure 3.
\begin{figure}[h!]
    \centering\includegraphics[scale=0.3]{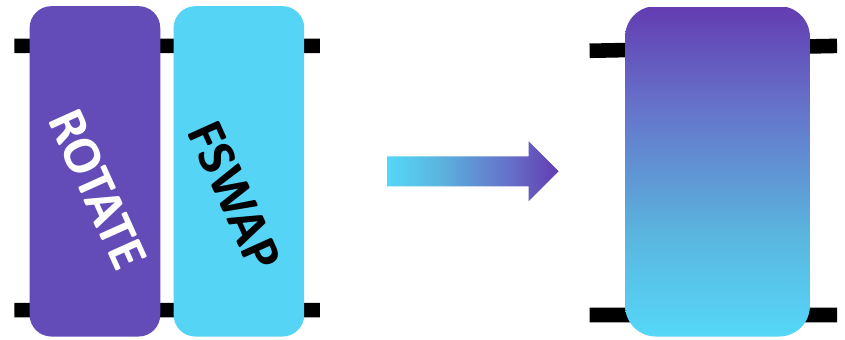}
    \caption{Quantum circuit representation for the combined effect of population transfer and fermionic swap operator. The resultant two-qubit operator also follows the YBE relation. }
\label{fig:merge_fswap_rotation}
\end{figure}
Therefore, combining the action of $\hat{h}_2$ with the FSWAP operator, one can design a circuit for a time step $\delta t$ such that each fermion is able to interact with the rest; see Figure 4. 
\begin{figure}[h!]
    \centering
    \includegraphics[width=\columnwidth]{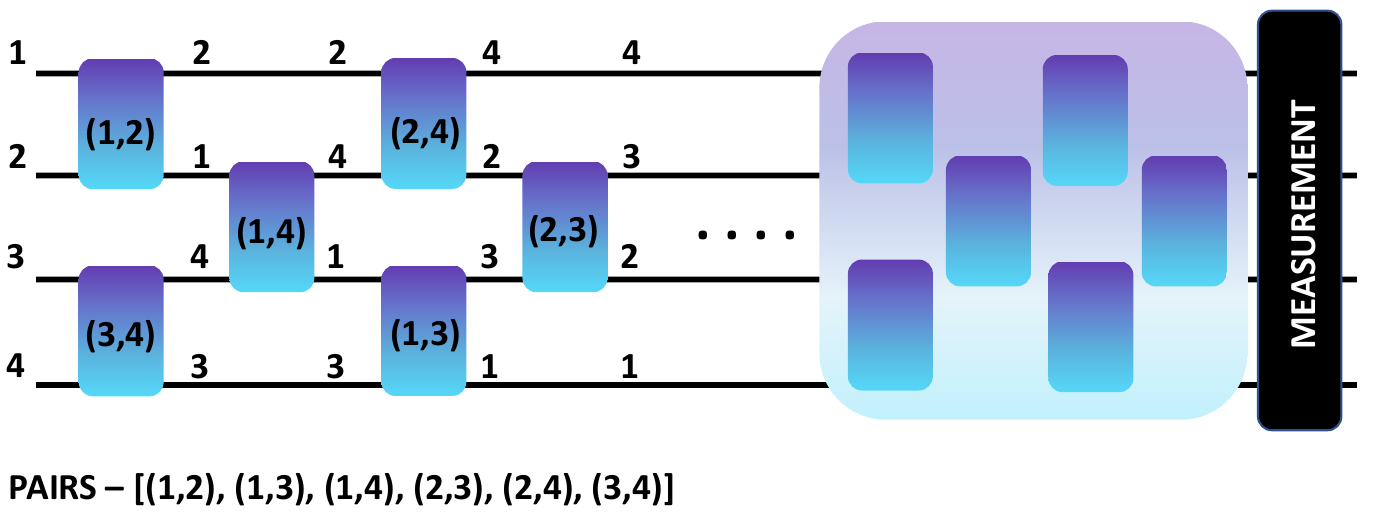}
    \caption{Combination of interaction and FSWAP operator in a layer form allows all the orbitals to interact. The label on the wire shows the identity before and after the action of gate. Two alternative layers cover all possible interactions for one Trotter step.}
    \label{fig:4qubit_qcircuit}
\end{figure}

This quantum circuit illustrates the design with four fermions labeled as (1,2,3,4). There are six possible interactions, and all the possible pairs that can be written for them are [(1,2),(1,3),(1,4),(2,3),(2,4),(3,4)]. The label on the wire shows the identity before and after the action of gate (composed of both the interaction and FSWAP). By the end of two alternative layers all possible interactions are covered for one Trotter step, as shown in Figure \ref{fig:4qubit_qcircuit}. For time dynamics with $n$ Trotter steps, one has to repeat the unit $n$ times. %\textcolor{red}{SKG: earlier, $n$ was used for number of orbitals.} 

We now prove that the combined action above also follows the YBE,  and this will allow one to compress the arbitrary size circuit to finite depth. The quantum gate representations for interaction matrix($G_{ij} = 1$) and FSWAP are
\begin{equation}
\begin{aligned}
&e^{-it(a_{i}^{\dagger}a_{i+1}+a_{i+1}^{\dagger}a_{i})} =  \\
    &
    \Qcircuit @C=0.5em @R=.7em @!R{
    & \gate{R_x(\pi/2)} &  \ctrl{1} & \gate{R_x(t/2)} & \ctrl{1} &  \gate{R_x(-\pi/2)}& \qw
    \\ 
    & \gate{R_x(\pi/2)} & \targ    & \gate{R_z(t/2)} & \targ  & \gate{R_x(-\pi/2)}& \qw
    } 
\end{aligned}
\end{equation}

and 
\begin{equation}
    \mbox{FSWAP} = \Qcircuit @C=0.5em @R=.7em @!R{
    & \gate{R_x(\pi/2)} &  \ctrl{1} & \gate{R_x(\pi/2)} & \ctrl{1} &  \gate{R_x(-\pi/2)}& \qw
    \\ 
    & \gate{R_x(\pi/2)} & \targ    & \gate{R_z(\pi/2)} & \targ  & \gate{R_x(-\pi/2)}& \qw .
    } 
\end{equation}
Combining both  actions, we have 
\begin{equation}
\begin{aligned}
&\mbox{FSWAP} \times e^{-it(a_{i}^{\dagger}a_{i+1}+a_{i+1}^{\dagger}a_{i})}  = \\ 
       &
    \Qcircuit @C=0.5em @R=.7em @!R{
    & \gate{R_x(\pi/2)} &  \ctrl{1} & \gate{R_x(t/2+\pi/2)} & \ctrl{1} &  \gate{R_x(-\pi/2)}& \qw
    \\ 
    & \gate{R_x(\pi/2)} & \targ    & \gate{R_z(t/2+\pi/2)} & \targ  & \gate{R_x(-\pi/2)}& \qw .
    } 
\end{aligned}
\end{equation}
Therefore, we have established that we require $n/2$ alternative layers for $n$ qubits to perform one time step and that the 2-qubit action combined with FSWAP follows the YBE. 
\begin{figure}[h!]
    \centering\includegraphics[width=\columnwidth]{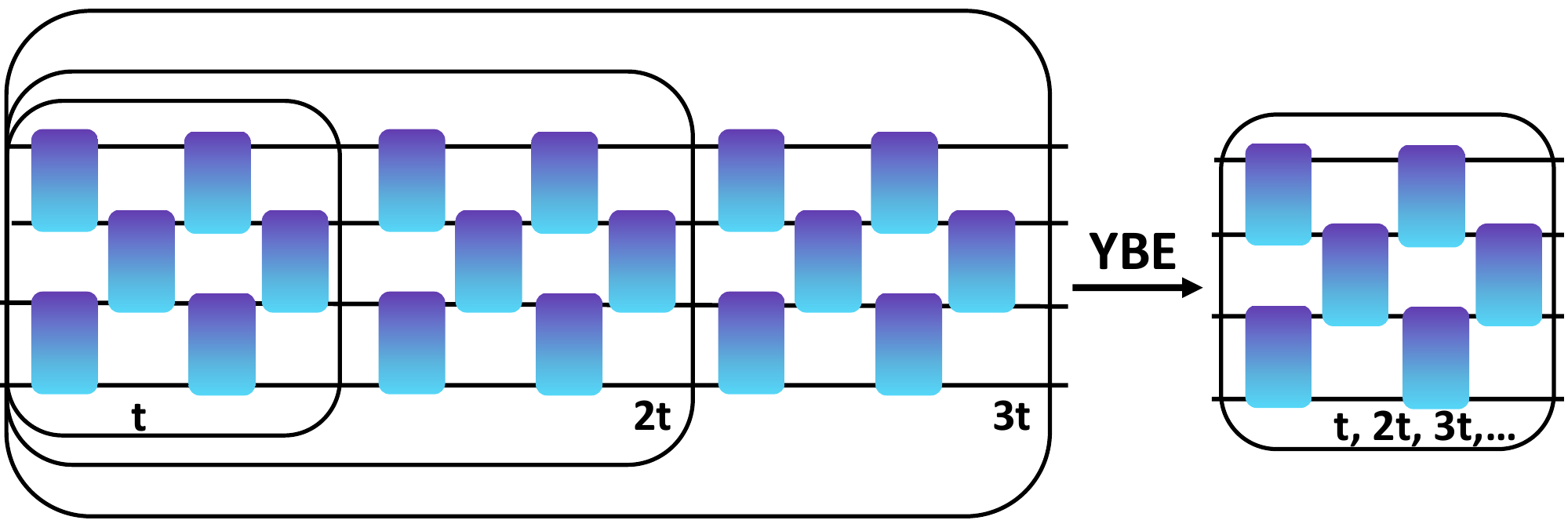}
    \caption{Circuit design for 4 qubits showing each time step. Each block on the left can be compressed to the right circuit.}
    \label{fig:circuit-design}
\end{figure}
The  circuit shown in Figure 5 follows the YBE, and each block can be compressed to a previous block. 

So far we have established that time dynamics of $n$ fermions with only the interaction matrix can be simulated with $n/2$ alternative layers efficiently. The leftover part is the number operator interaction, namely,  $\hat{h}_1$. Each element in $\hat{h}_1$ commutes with each other,  allowing us to write the time dynamics operator as a product of exponentials for each element. 
\begin{equation}
    e^{-it(G_{11}\hat{n}_{1}+G_{22}\hat{n}_{2}+...)} = e^{-it(G_{11}\hat{n}_{1})}\times e^{-it(G_{22}\hat{n}_{2})} \times .... 
\end{equation}
On further analysis of the individual elements we also find
\begin{equation}
e^{-it(G_{ii}\hat{n}_{i})} = P(tG_{ii}),
\end{equation}
\begin{figure}[h!]
    \centering
    \includegraphics[scale=0.4]{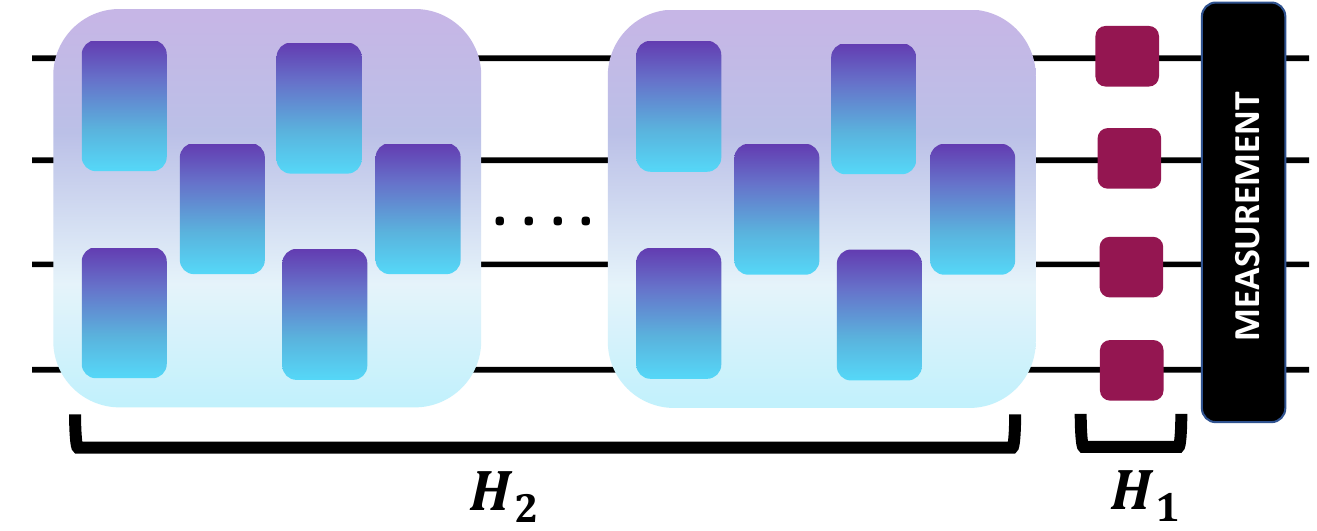}
    \caption{Final circuit for TDHF.}
    \label{fig:final_circuit}
\end{figure}
where $P(\phi_{i})$ is a phase gate on the $i$th qubit. The  phase gate adds a phase to a quantum state and can be acted on the qubits by the end of action by $\hat{h}_2$. 
Therefore the final quantum circuit  relevant for TDHF is shown in Figure 6.
%\ref{fig:final_circuit}. 
The whole circuit is composed of two qubit gates that follow the YBE relation and one qubit rotation that acts at the end. 
Both components are compressible individually and allow one to achieve a linear depth circuit for each evolution step. 

The workflow for the time dynamics of the Hartree--Fock equation on a quantum simulator is as follows.
\begin{enumerate}
    \item Prepare the initial Hartree--Fock state.
    \item Generate the quantum circuit for $\hat{h}_1$ and $\hat{h}_2$ using the Fock matrix and dipole moment interaction with the electric field for the first time step.
    \item Perform the time dynamics for the first step, and measure the energy.
    \item Use the Pauli Z measurement of each qubit to determine the new density matrix using Eq.~\ref{eq:density}. 
    \item Convert the density matrix, which is in molecular basis, back to the density matrix in atomic basis by using the coefficient matrix from the Hartree--Fock ground state.. 
    
    \item Use the previous density matrix to compute the new Fock matrix and generate the quantum circuit forthe  second time step.
    \item Perform the compression on actions of $\hat{h}_1$ using single-qubit compression and for $\hat{h}_2$ using the YBE compression technique. 
    \item Perform the time dynamics with the compressed circuit, and repeat steps 4--6 until the whole dynamics is finished.
\end{enumerate}
One of the major drawbacks of this formalism is the construction of the modified Fock matrix at each step. As the Fock matrix itself evolves with time, it forces us to include that perturbation. We do so by measuring the changes in population of molecular spin orbitals and construct the density matrix using Eq.~(12). Therefore, it forces us to compute the modified Fock matrix classically. 
This formalism will be true for any time-dependent Hamiltonian and will act as a bottleneck for the simulation of quantum-time dynamics. 
%\textcolor{red}{SKG: (a) I don't understand step 4 -- fluctuation in density matrix? (b) Since only a few qubits are involved could an Appendix give the qubit form of H1 and H2 and/or the compressed circuits? You state in words later that it sounds like there might only be 12 gates but it would be nice to be explicit.  BTW if so few gates then is there the possibility of actual implmentation on a quantum computer or is there some snag?}
%\textcolor{red}{SKG: need to state more explicitly where the classical computer computations are being done above and that this is a hybrid method. }

\section{Results}
Here we show, in  simulation, how our hybrid classical/quantum TDHF applies to the problem of H$_{2}$ interacting with an electric field. The electric field envelope $|E(t)|$ is linearly increased with time to a maximum value of $|E_{max}|$ at the end of the first cycle; it remains $|E_{max}|$ for one cycle and then decreases to zero by the end of the next cycle.
%\textcolor{red}{SKG: use of "e" in theory section and "E" here for electric field. Schlegal had some technical differences but for us maybe just
%use "E" everywhere for simplicity?}
\begin{equation}
    \begin{aligned}
        E(t) = (\omega t/2\pi)\mbox{sin}(\omega t)E_{max} & \mbox{ 
  for } & 0\leq t \leq 2\pi/\omega \\
        E(t) = \mbox{sin}(\omega t)E_{max} & \mbox{   for } & 2\pi/\omega\leq t \leq 4\pi/\omega\\
        E(t) = (3-\omega t/2\pi)\mbox{sin}(\omega t)E_{max} & \mbox{    for } & 4\pi/\omega\leq t \leq 6\pi/\omega\\
        E(t) = 0 & \mbox{    for } & t\leq 0 \mbox{  and  } t\geq 6\pi/\omega
    \end{aligned}
\end{equation}
\begin{figure}[h!]
    \centering\includegraphics[width=\linewidth]{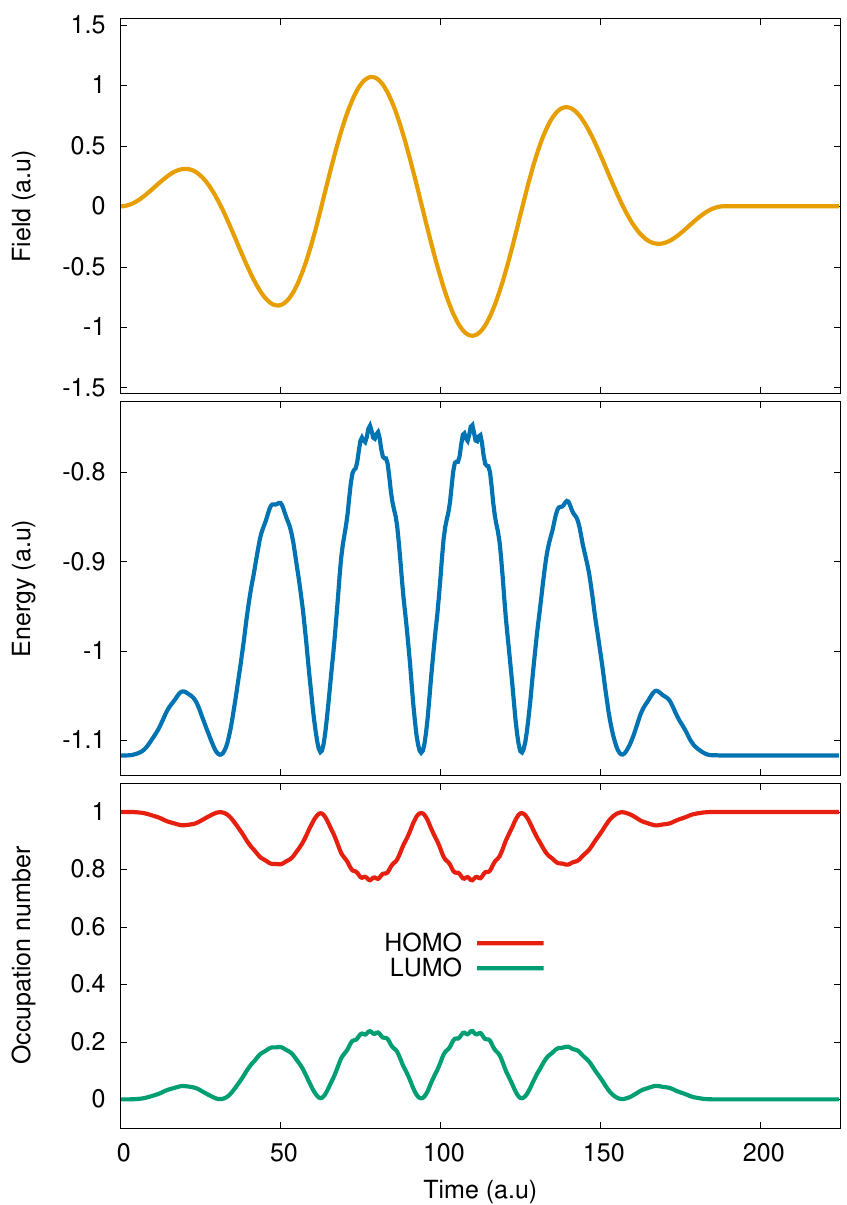}
    \caption{TDHF simulation for H$_{2}$ in an oscillating electric field ($E_{max}$ = 1.07 a.u. (1.72 $\times$ 10$^{14}$ W cm$^{-2}$) and $\omega$ = 0.10 a.u. (456 nm)) using STO-3G basis. (a) Electric field profile. (b) Total energy evolution with the oscillating field. (c) Electron populations of the HOMO and LUMO.}
    \label{fig:h2_sto3g}
\end{figure}

%\textcolor{red}{SKG: I don't undertstand the  above -- is there a missing cos(omega t)in places on the rhs?}

Figure  7 shows the response of an H$_{2}$ 
molecule at the equilibrium geometry of 
$R_e = 0.734 \AA$ 
(HF/STO-3G). 

Figure  7c shows the electron population evolution for both the HOMO and LUMOs. This simulation was performed on a quantum simulator (PennyLane \textcolor{red}{[ref]}) with the parameterized circuit using Eq. ~(32).
There are four orbitals that require four qubits for this two-electron problem. The time dynamics from $\hat{h}_1$ requires only 4 CNOTS and 8 one-qubit rotation gates. On the other hand, the action of $\hat{h}_2$ requires only 4 one-qubit rotation gates. 
The whole evolution of system always contains only finite-depth gates due to compression mediated by YBE relations. 
%\textcolor{red}{SKG:Previous sentence not understandable.}
The parameters for the gate operations were obtained by using PySCF~\cite{sun2018pyscf} for the H$_{2}$ molecule ($R_e = $0.734 a.u). %\textcolor{red}{SKG: do not understand "by RHF using"}

%\textcolor{red}{SKG:need to give much more detail -- do we have just two qubits, one for each of the two electrons/orbitals, etc.? I guess with the field on they somehow morph into excited orbitals and then, as we discussed, relax down when the field is off? Wouldn't hurt to discuss this stuff.}

\begin{figure}[!ht]
    \centering\includegraphics[width=\linewidth]{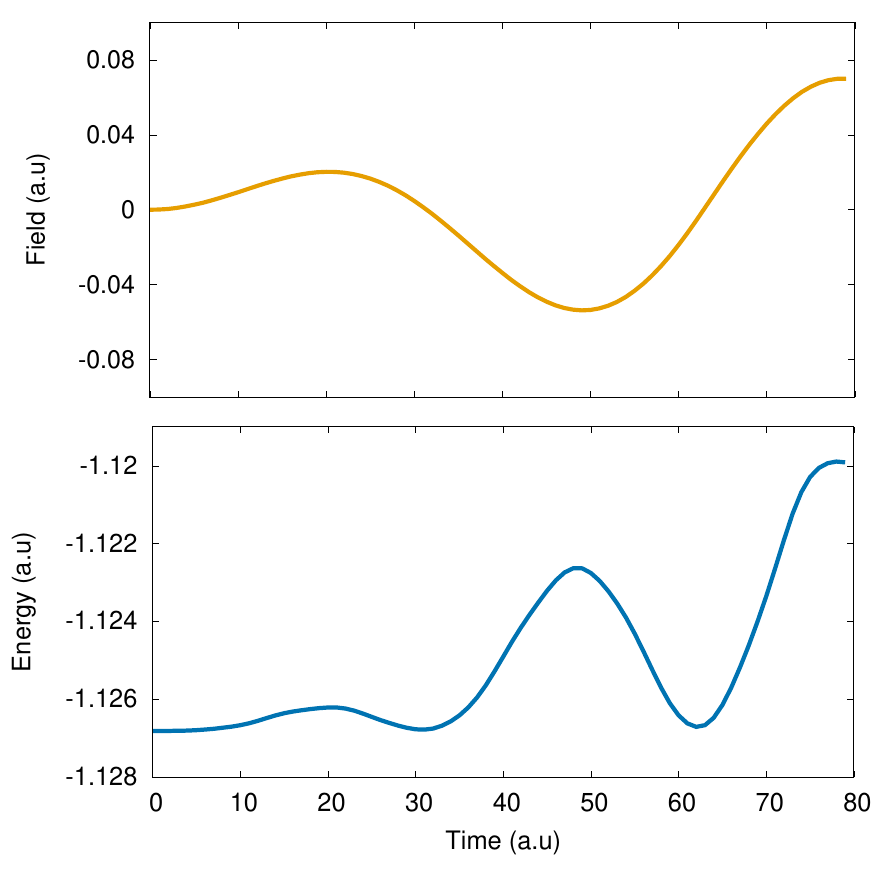}
    \caption{TDHF simulation for H$_{2}$ in an oscillating electric field ($E_{max}$ = 0.07 a.u. (1.72 $\times$ 10$^{14}$ W cm$^{-2}$) and $\omega$ = 0.10 a.u. (456 nm)) using 6-31G basis. (a) Electric field profile. (b) Total energy evolution with the oscillating field.}
    \label{fig:h2_631G}
\end{figure}
To show the scaling of the technique developed in this article, we also performed the same simulation for larger basis. Figure 8 %\ref{fig:h2_631G} -- I spelled out the ref. because Figure should be spelled out when starting a sentence
shows the H$_{2}$ molecule in the same electric field as described above but with a 6-31G basis set. As one can see, the dynamics is  similar to that of the minimal basis set. We compared these results with the canonical TDHF simulation and proved the accuracy of the simulations. 

\section{Conclusion}
We have shown that YBE symmetry mediates the compression of the quantum circuit for the time-dependent Hartree--Fock method. The depth of the quantum circuit scales linearly with the number of spin orbitals for unrestricted Hartree--Fock and the number of molecular orbitals for restricted Hartree--Fock levels. For each time step the information from the previous step is required, thus forcing a measurement after each time step to construct the updated Fock matrix. We demonstrated the technique's efficacy on a quantum simulator. The algorithm is a hybrid classical-quantum algorithm: It uses quantum device to perform the time evolution, while the parameters required to do it are computed classically. It does not have a quantum advantage, but it sets a foundation for the use of the YBE-mediated compression technique in quantum chemistry. %\textcolor{red}{SKG: Again, it is only implicit that classical computations are needed since "forces a measurement" does not go further in explanation.}

% SKG commented this out since everyone says it.
%This work paves the path to demonstrate quantum advantage on NISQ devices. Time-dynamics being a prime-candidate which generally have large-depth quantum circuits. YBE symmetry allow us to obtain small depth circuit without any compromise on accuracy. 

\section*{Acknowledgment}
This material is based upon work supported by the U.S. Department of Energy, Office of Science, National Quantum Information Science Research Centers. Y.A. acknowledges 
support from the U.S. Department of Energy, Ofﬁce of Science, under contract DE-AC02-06CH11357 at Argonne National Laboratory. 
Y.A. and S.G.  acknowledge Arthur Izmaylov and James Whitfield for insightful discussion.

\section*{Appendix A}
In this section we explain the general proof for circuit for fully connected graph of fermions. 

\begin{theorem}
Let $G:(N,E)$ be a fully connected graph with $V$ the number of nodes.  $E=N(N-1)/2$ is the number of edges representing the interactions between nodes. Then $E$ can be mapped to a layered circuit structure shown in Figure~\ref{fig:layered-circuit},
\begin{figure}[h!]
    \centering\includegraphics[scale =0.4]{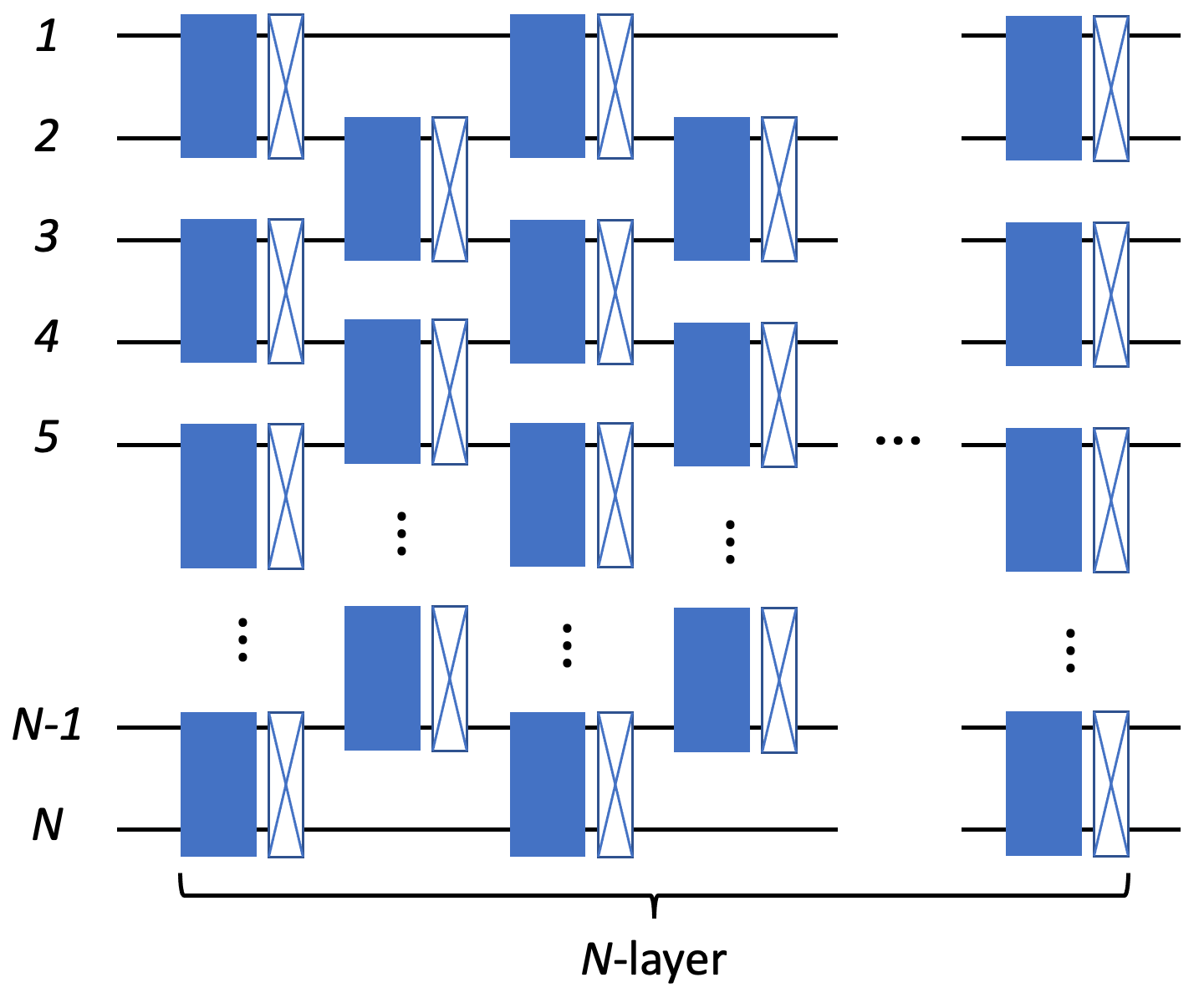}
\label{fig:layered-circuit}
\end{figure}
where each edge in the graph $G$ is represented by a two-qubit rotation that is followed by a fermion swap operation. 
\end{theorem}

The proof of \textbf{Theorem 1} can be done in two steps. First, the edges in a fully connected graph $G$ can be represented as  in Table \ref{tab:connection}.

\begin{table}[h!]
    \centering
    \begin{tabular}{llllll}
        (1,2) & (1,3) & (1,4) & $\cdots$ & (1,N-1) & (1,N) \\
        (2,3) & (2,4) & (2,5) & $\cdots$ & (2,N) &  \\
        $\vdots$ & $\vdots$ & $\vdots$ & & & \\
        (N-3,N-2) & (N-3,N-1) & (N-3,N) &&& \\
        (N-2,N-1) & (N-2,N) &&&& \\
        (N-1,N) & &&& & \\
    \end{tabular}
    \caption{All the connections in a fully connected graph $G:(N,E)$.}
    \label{tab:connection}
\end{table}

Suppose we can use a two-qubit rotation to represent the connection between two nodes; that is,  for $(p,q)$ we can have a two-qubit rotation to link qubits $p$ and $q$. If $p$ and $q$ are next two each other, such a two-qubit rotation can be applied straightforwardly. Otherwise we can keep applying fermion swap operations to enforce the adjacency of $p$ and $q$. Now look at the first row. Such a two-qubit rotation followed by a fermion swap can be applied continuously to get all the connections in this row to be encoded. The resulting circuit will be the following.
\begin{figure}[h!]
    \centering\includegraphics[scale=0.4]{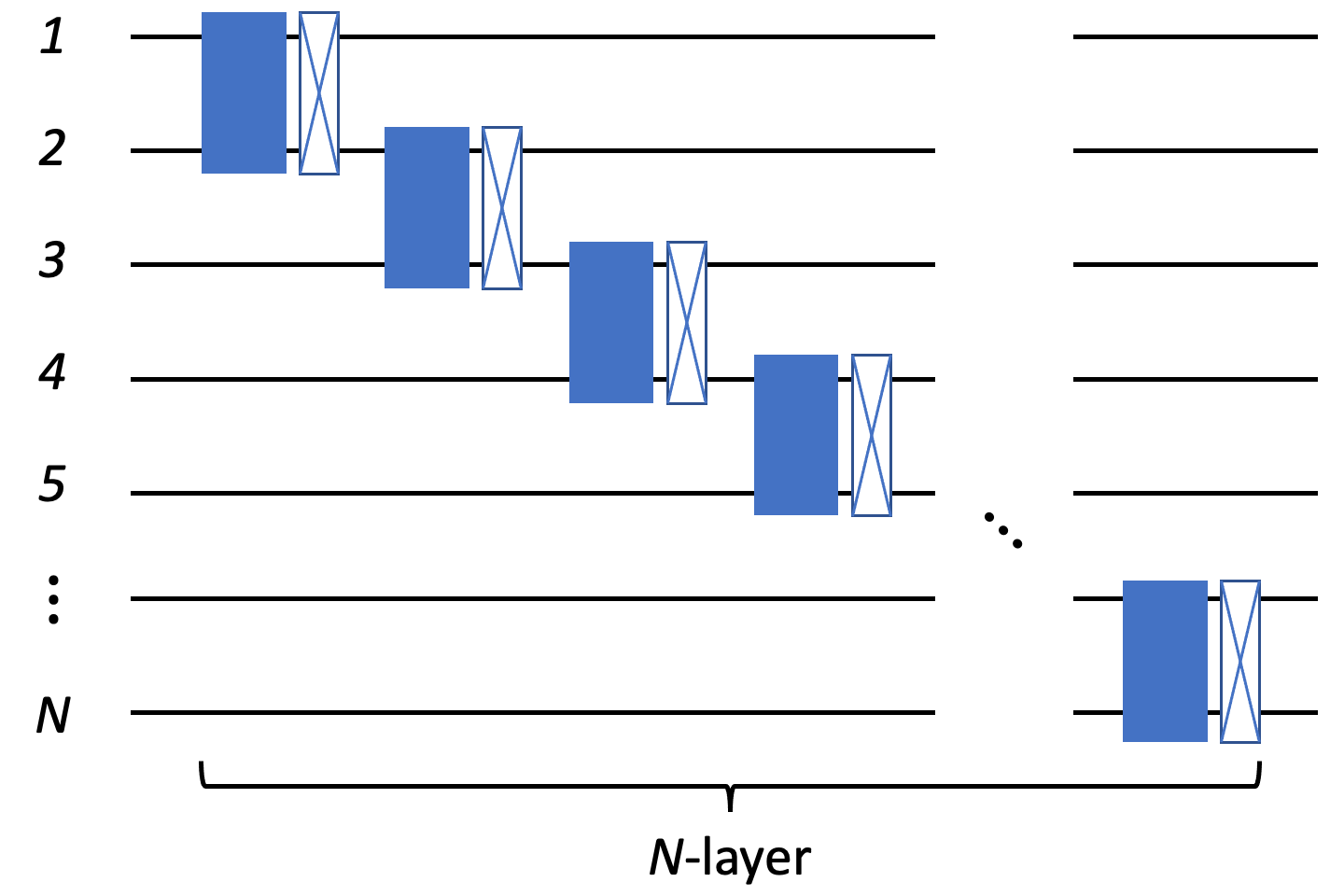}
\label{fig:layered-circuit2}
\end{figure}

We can then proceed to the second row, the third row, $\cdots$, all the way to the $(N-1)$th row to end up with a triangular circuit structure as shown below.
\begin{figure}[h!]
    \centering\includegraphics[width=\linewidth]{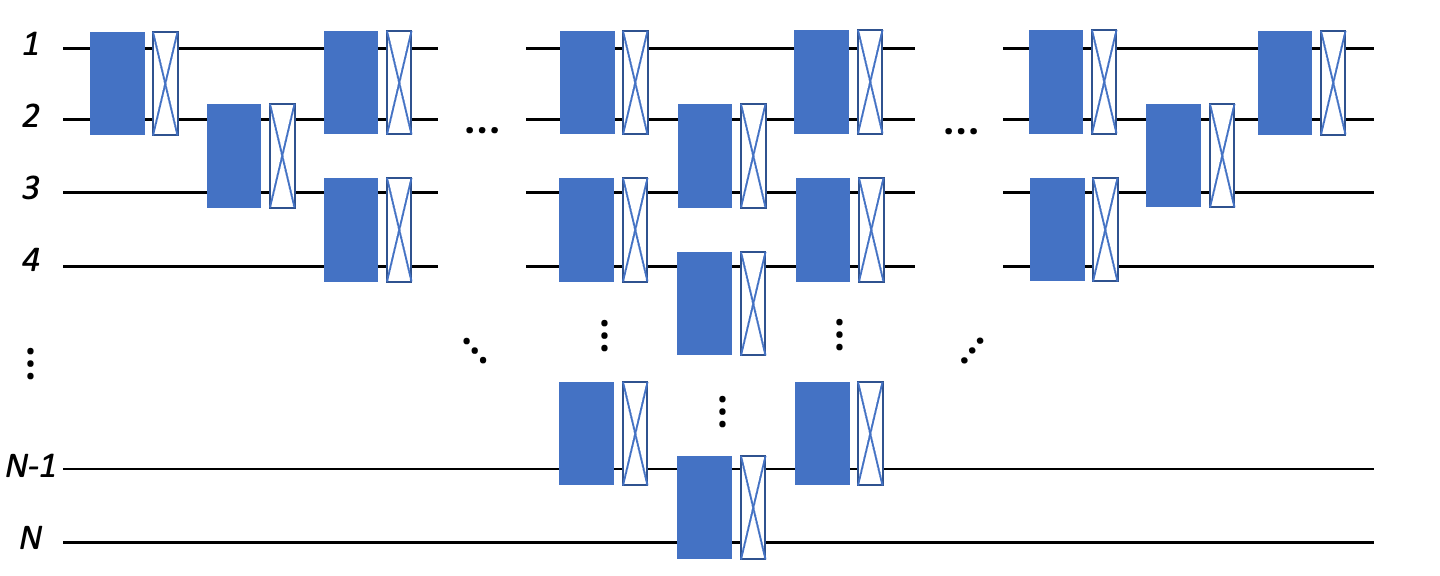}
\label{fig:layered-circuit3}
\end{figure}

Now, if the building block, namely, a two-qubit rotation followed by a fermion swap, satisfies the Yang--Baxter-like equation as show in Figure \ref{fig:YBE_rep}, then the triangular circuit structure can be equivalently transformed to a layered circuit structure through limited number of YBE-like transformations. A possible strategy would be starting from the rightmost two-qubit gate in the above circuit and moving gradually to the bottom left of the circuit through consecutive YBE-like transformations as shown below.
\begin{figure}[h!]
    \centering\includegraphics[width=\linewidth]{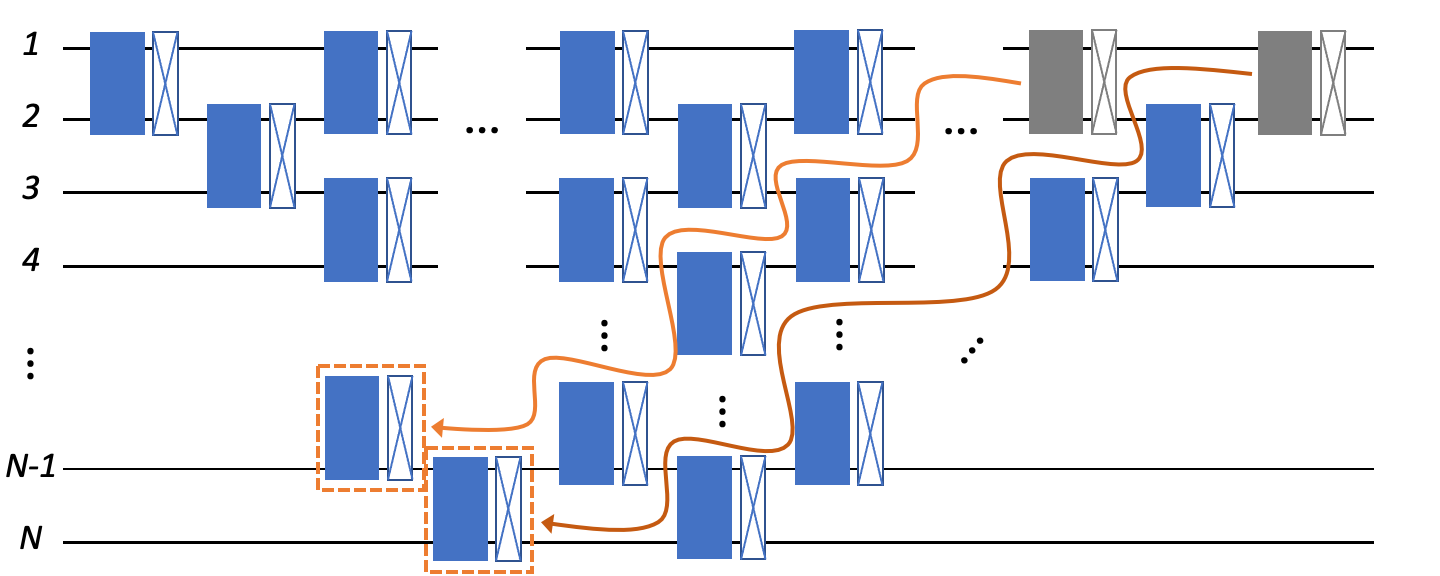}
\label{fig:layered-circuit4}
\end{figure}

Eventually, through a total number of $\mathcal{O}(N^3)$ YBE-like transformations,  one will get the layered circuit structure.

\bibliography{references}

%\begin{widetext}
%\vfill
%\footnotesize

%\framebox{\parbox{\textwidth}{
%The submitted manuscript has been created by UChicago Argonne, LLC, Operator of Argonne National Laboratory (``Argonne''). Argonne, a U.S.\ Department of Energy Office of Science laboratory, operated under Contract No.\ DE-AC02-06CH11357 and Pacific Northwest National Laboratory (PNNL), a U.S.\ Department of Energy Office of Science laboratory, operated by Battelle Memorial Institute for the United States Department of Energy under Contract No DE-AC05-76RL1830. The U.S.\ Government retains for itself, and others acting on its behalf, a paid-up nonexclusive, irrevocable worldwide license in said article to reproduce, prepare derivative works, distribute copies to the public, and perform publicly and display publicly, by or on behalf of the Government.  The Department of Energy will provide public access to these results of federally sponsored research in accordance with the DOE Public Access Plan. 
%http://energy.gov/downloads/doe-public-access-plan}}.
%\end{widetext}
\end{document}